\documentclass[fleqn,twoside]{article}
\usepackage{espcrc2}

\usepackage{graphicx}

\def\BB{$\beta\beta$ }
\def\BBz{$\beta\beta[0\nu]$ }
\def\BBt{$\beta\beta[2\nu]$ }

\def\amee{$\vert\langle m_{ee} \rangle\vert$~}
\def\ca{$\sim$}

\title{Probing low energy and mass scales}

\author{Oliviero\ Cremonesi\address{INFN Sez. di Milano Bicocca and Dipartimento
    di Fisica ``G.Occhialini'', Universit\'a degli studi di Milano Bicocca, Piazza della Scienza 3, I-20126
    Milano,Italy} and Alessandro\ Melchiorri\address{Dipartimento di Fisica and Sezione INFN, Universit\`a degli Studi di Roma ``La Sapienza'', P.le Aldo Moro 5, 00185, Rome,
Italy}}

\begin{document}
%\date{{\today}}
\begin{abstract}
Low energy neutrino processes are ideal probes for new Physics beyond
 the Standard Model. Cosmological observations and rare nuclear
 processes can test neutrino mass scales and give definite answers to
 unsolved basic questions like the Majorana/Dirac nature of the
 neutrino.
% We summarize the experimental and theoretical efforts discussed in the  ``Probing low energy and mass scales'' session and give a short overview of the future perspectives, referring to the individual contributions for more details.
\end{abstract}

\maketitle

%%%%%%%%%%%%%%%%%%%%%%%%%%%%%%%%%%%%%%%%%%%%%%%%%%%%%%%%%%%%%%%%%%%%%%
%%%% Section I %%%%%%%%%%%%%%%%%%%%%%%%%%%%%%%%%%%%%%%%%%%%%%%%%%%%%%%
%%%%%%%%%%%%%%%%%%%%%%%%%%%%%%%%%%%%%%%%%%%%%%%%%%%%%%%%%%%%%%%%%%%%%%

\section{Introduction}
Well proved by a number of experiments carried out during the last ten years, the existence of neutrino oscillations has stimulated renewed interest in the study of neutrino properties. Neutrino oscillation results have shown that neutrinos mix and have finite masses; they have thus provided us with the first clear evidence of phenomena beyond the reach of Standard Model (SM). 
Unlike quark mixings, neutrino mixings are large, although the reason for this is not yet fully understood. On the other hand neutrino masses are much smaller than those of the charged leptons and their pattern (or absolute scale) is still unknown.  Neutrino oscillations in fact only depend on the absolute value of the difference of the squares of the neutrino masses and two possible hierarchies (or orderings) are then implied by current available data: the normal ($m_1 < m_2 \ll m_3$) and the inverted hierarchy ($m_3 \ll m_1 < m_2$).
These and other unsolved questions concerning neutrino properties are becoming subject of increasing interest and considered a unique tool to see what new Physics lies beyond SM predictions. This holds in particular for the Dirac/Majorana neutrino nature which represents one of the most important open questions in neutrino Physics. In the SM neutrinos are Dirac particles by construction (i.e. in order to conserve lepton number L). In the limit of vanishing masses however Lepton Number conservation can be equivalently stated in terms of neutrino helicity properties and the Majorana or Dirac descriptions for neutrino are equivalent (i.e. don't change the physical content of the theory). For finite neutrino masses however the two descriptions are no more equivalent and can give rise to different Physical scenarios (e.g. mass generation mechanisms).
 
Only experiments sensitive to m$_1$ can aim at solving the mass hierarchy problem. This is the case for kinematic measurements of the $\beta$ spectrum end-point, neutrinoless double beta decay (\BBz) and cosmological measurements. 
Intense efforts are therefore being spended to fix (measure or bound) the absolute scale of neutrino masses (as contrasted to the $\Delta m^2$ measurements of neutrino oscillation experiments) with various observations and experiments, characterized by different systematics and sensitivities. Cosmological observations and laboratory experiments actually probe different quantities which can be however compared within specific models. In the case of 3 Majorana neutrinos and upon a good control of the systematics, cosmological observations and \BBz experiments have the largest success potential. In the same theoretical framework, complementary informations can come also from oscillation experiments if U$_{e3}$ is large enough. Kinematical measurements seem presently characterized by a lower sensitivity although they represent the only model independent measurement of the neutrino mass scale. On the other hand \BBz experiments are the only presently available tool to probe the Dirac/Majorana nature of the neutrino.

It should be stressed that the 3 Majorana neutrino model is the most likely but not the unique framework for the measurements discussed above. Different extensions of the SM could equally account for such observations. Moreover there is no possibility to account for the LSND result within the the 3 Majorana neutrino model.

The present status and future perspectives of the esperimental (cosmological observations and laboratory measurements) and theoretical efforts in the framework of neutrino masses were discussed in dedicated talks of the present session and in variuos talks of the general sessions. A general overview from the point of view of terrestrial and sky observations will be given in the next sections. A possible scenario for future developments in the forthcoming years will be summarized in the last section.
 
\section{Laboratory measurements}
Neutrino absolute mass scale and nature are two of the most relevant open questions in neutrino Physics addressed by next generation laboratory experiments.
All proposed experiments are based on the precise study of proper nuclear processes: single beta decay for the direct measurement of the electron antineutrino mass and neutrinoless double beta decay as a unique probe of the basic neutrino properties.

\subsection{Neutrinoless $\beta\beta$ decay}
Having no conserved charge except for the Lepton number L, neutrinos are the only truly ``neutral'' fermions characterized by the possibility of being their own anti-particles. If, as predicted in many extensions of the SM, L is in fact not conserved, neutrinos could be the unique fermions with a Majorana nature. The implications of Majorana or Dirac massive neutrinos are completely different and the question of their nature is therefore an essential building block for every possible extension of the Standard Model. The only practical method to attack this problem is the study of neutrinoless double beta decay~\cite{cremo04}, a rare nuclear process in which a nucleus (A, Z) would turn into an isobar (A,Z+2) by transforming two neutrons into protons while emitting two electrons: (A, Z) $\to$ (A, Z+2) + 2e$^-$ . This differs from "normal" double-beta decay or \BBt (second order process of the weak interaction), which is rare but has been detected: (A, Z) $\to$ (A, Z+2) + 2e$^-$ + 2$\nu$. \BBz can occur through different processes but all of them require that the neutrino is a Majorana particle. Moreover, when induced by the exchange of a massive Majorana neutrino, the \BBz width is proportional to an effective parameter (neutrino mass) \amee=$\Sigma_{k}U_{ek}^2 m_k$ through the phase space factor G$^{0\nu}$ and nuclear matrix element M$^{0\nu}$
%...................................................................
\begin{equation}\label{Tau}
(\tau)^{-1} = G^{0\nu}\vert M^{0\nu} \vert^2 \langle m_{\nu} \rangle^2/m_e^2
\end{equation}
%...................................................................

\BBz observation would allow to obtain therefore relevant informations on the neutrino absolute mass scale. 
Large uncertainties in the nuclear matrix elements calculations imply serious difficulties in the comparison between experiments on different \BBz active isotopes and represent the most relevant source of systematics in the evaluation of \amee. The subject was discussed in the talk of Fedor Simkovic who described a new calculation method, based on the normalization of the relevant QRPA parameters to the \BBt results, characterized by a better stability with respect to previous QRPA calculations. \BBt results are unfortunately not available for all \BBz active isotopes and the method is not yet fully accepted by the whole community involved in M$^{0\nu}$ calculations. More efforts in this field are therefore strongly encouraged and experimental analyses on different active \BBz isotopes are required.

Besides the general motivations discussed above, \BBz is presently a hot and debated subject for two more reasons: i) the claim for a positive effect in the Heidelberg-Moscow data on $^{76}$Ge; ii) in the inverted hierarchy scenario, present neutrino oscillation results suggest a \amee value of the order of 50 meV, in the reach of next generation experiments. 

Two presently running \BBz experiments have the potential to confirm the $^{76}$Ge  claim and probe the inverted hierarchy region of neutrino masses: CUORICINO~\cite{CINO} at LNGS and NEMO3~\cite{NEMO3} in the Frejus tunnel.

Based on the bolometric technique, CUORICINO exploits a fully calorimetric approach and was introduced in the talk of Maura Pavan. A limit of 2.4 $\times$ 10$^{24}$ y on the half-lifetime for \BBz of $^{130}$Te was presented, corresponding to a limit on the effective neutrino mass in the range 0.18-1.05 eV.
With a total mass of nearly 41 kg of natural TeO$_2$ and a satisfactory duty cycle, CUORICINO is running since 2003 in Hall A of the Gran Sasso underground Laboratory. The achieved background level is 0.18 $\pm$ 0.01 c/keV/kg/y corresponding to a 5y sensitivity of 8.7 $\times$ 10$^{24}$ y on the $^{130}$Te \BBz half-lifetime.

Characterized by a similar sensitivity but a completely different approach, NEMO3 was discussed by Ladislav Vala. Advantaged by a well known technology, this tracking+calorimetric
detector can study different nuclei simultaneously and identify the \BB events by reconstructing the electron tracks. NEMO3 is installed in the Modane underground laboratory and has produced excellent results on the \BBt decay of various nuclei: $^{100}$Mo, $^{82}$Se, $^{150}$Nd, $^{116}$Cd and $^{96}$Zr. An upper limit of 5.8 $\times$ 10$^{23}$ y on the half-lifetime for \BBz of $^{100}$Mo was presented, corresponding to a limit on the effective neutrino mass in the range 0.6-0.9 eV. The projected 5 y sensitivity on $^{100}$Mo \BBz halflife is 4.0 $\times$ 10$^{23}$ y corresponding to a \amee 0.2-0.35 range. 
Both CUORICINO and NEMO3 have the sensitivity to confirm the $^{76}$Ge claim. A disproof is however impossible for both of them because of the large uncertainties in the nuclear matrix elements calculations. Different would be the case for $^{76}$Ge experiments for which a direct comparison would be possible\cite{GERDA}.
A defnite answer to the $^{76}$Ge claim should be in any case possible with the upcoming next generation experiments, characterized by improved masses (\ca 1 ton) and background levels (1-10 counts/keV/ton/y). Main goal of these experiments is to probe the inverted hierarchy of the Majorana neutrino mass scale. Sensitivities better than  \ca 50 meV on \amee are therefore foreseen. In most cases they are extensions of previous experiments exploiting the most successful techniques. Variuos 1 ton experiments have been so far proposed and some of them are already in construction. A general introduction to the USA efforts (mainly Majorana~\cite{Majo} on $^{76}$Ge and EXO~\cite{EXO} on $^{136}$Xe) was given by Frank Avignone. The bolometric CUORE~\cite{CUORE} experiment on $^{130}$Te, of which CUORICINO is a successful demonstrator, is in construction at LNGS. Its completion is foreseen within the end of 2010 and was presented by Maura Pavan. Alessandro Bettini introduced GERDA~\cite{GERDA}, also in construction at  LNGS, whose first phase (with just 20 kg of isotopically enriched Ge) would allow to definitely test the $^{76}$Ge claim within 2010. Finally, SUPERNEMO~\cite{SNEMO} (extention of NEMO3) was disussed by V.~Lala.

\subsection{Single $\beta$ decay}
The precise measurement of the electron spectrum in $\beta$-decays is the only laboratory technique for the direct measurement of a small neutrino mass, without additional assumptions on the character of the neutrino. The neutrino mass (or an upper limit to it) is inferred from the shape of the energy spectrum near its kinematical end point. The shape of the spectrum near the kinematical end point depends on the masses of all three mass states and the measurement is sensitive to an effective mass m$_\beta$=$\sqrt{\Sigma_k \vert U_{ek}\vert^2 m_k^2}$ . The relative number of events occurring in an interval of kinetic energy $\Delta$T near the end-point is proportional to ($\Delta$T/Q)$^3$ so that very low Q-values are required. Even exploiting the lowest Q-value $\beta$-processes available in nature ($^{187}$Re and $^{3}$H with 2.5 and 18.6 keV respectively) it is however difficult to obtain sensitivities much better than \ca 1 eV. These experiments can therefore probe only quasi-degenerate neutrino masses (direct hierarchy). They represent however the only model independent approach to neutrino  masses and are still therefore of key importance. \BBz experiments assume in fact the Majorana nature of the neutrino while cosmological observations depend on a number of astrophysical assumptions. Moreover, the projected sensitivities are in the range of the $^{76}$Ge \BBz claim, implying a substantial discovery potential of the upcoming experiments in case this claim would be confirmed.

So far, the best results have been obtained with magnetic spectrometers (MAINZ~\cite{Mainz} and TROITSK~\cite{Troitsk}) which have reached sensitivities of the order of 2 eV. KATRIN~\cite{KATRIN}, the only tritium experiment presently foreseen, has been discussed in the general talk of C.~Weinheimer. Its projected sensitivity is 0.2 eV and will represent probably the end of the $^{3}$H  program. The KATRIN setup preparation is continuing without relevant delays and data taking should reasonably start in 2009.

Bolometric techniques offer a true calorimetric approach characterized by completely different (and probably safer) systematic uncertainties. A general presentation on the subject was given by Monica Sisti. Present sensitivities are however in the range of few eV\cite{mibeta,manu2} while the proposed large scale extensions (MARE~\cite{mare}) should dare to reach the 0.1 eV region in the forthcoming decade. On the other hand, would this approach be successful, its extension would be strightforward. 

\section{Cosmology}
Cosmological observations have started to provide valuable upper limits 
on absolute neutrino masses (see, e.g., the reviews \cite{Barg,Dolg}), 
competitive with those from laboratory experiments. 
In particular, the combined analysis of
high-precision data from Cosmic Microwave Background (CMB)
anisotropies and Large Scale Structures (LSS) has already reached a
sensitivity of $O(\mathrm{eV})$ (see, e.g., \cite{Be03,Tg04,Laha})
for the sum of the neutrino masses $\Sigma$,
%...................................................................
\begin{equation}\label{Sigma}
\Sigma = m_1+m_2+m_3\ .
\end{equation}
%...................................................................
We recall that the total neutrino energy density in our Universe,
$\Omega_{\nu}h^2$ (where $h$ is the Hubble constant normalized to
$H_0=100$ km~s$^{-1}$~Mpc$^{-1}$) is related to $\Sigma$ by the
well-known relation $\Omega_{\nu}h^2=\Sigma / (93.2 \mathrm{\ eV})$
\cite{PDG4}, and plays an essential role in theories of structure
formation. It can thus leave key signatures in LSS data 
(see, eg.,\cite{Hu98}) and, to a lesser extent, in CMB data 
(see, e.g.,\cite{Ma95}). Very recently, it has also been shown that accurate
Lyman-$\alpha$ (Ly$\alpha$) forest data \cite{Mc04}, taken at face
value, can improve the current CMB+LSS constraints on $\Sigma$ by a
factor of $\sim 3$, with important consequences on absolute neutrino
mass scenarios(\cite{Se04,fogli04,fogli06}). 

In our session, the importance of cosmology for constraining neutrino
masses and physics has been stressed in several talks.
In particular, the most recent upper limits on neutrino masses from
cosmology have been reported by Oystein Elgaroy (see \cite{elgaroy}). 
Those limits now reach the impressive sub-eV level, which is competitive with 
future terrestrial neutrino experiments. Elgaroy however pointed out 
some of the caveats that should be borne in mind when interpreting the
significance of these limits. 

Recent high redshift matter clustering data derived from
observations of Lyman-$\alpha$ forest clouds, have been
extremely powerful in constraining neutrino masses.
In his talk, Matteo Viel, explained in detail this new
and promising cosmological tool.
Lyman-$\alpha$ forest data allow to constrain the matter power spectrum 
from small scales of about 1 Mpc/h all the way to the horizon scale. 
The long lever arm and complementarity provided by such an analysis 
has previously led to a significant tightening of the constraints 
on the shape and the amplitude of the power spectrum of primordial 
density fluctuations. Viel then presented
 a combined analysis of the WMAP three year results with Lyman-$\alpha$
 forest data (see \cite{viel}). 

Gianpiero Mangano, presented new results from a recent analysis where 
the cosmological effects of interactions of neutrinos with cold Dark
Matter (DM) is investigated \cite{mangano}. 
This interaction produces diffusion-damped oscillations in the matter 
power spectrum, analogous to the acoustic oscillations in 
the baryon-photon fluid. 
It is therefore possible to obtain new and
independent bounds from galaxy surveys like the 
recent Sloan Digital Sky Survey on 
the corresponding opacity defined as the ratio of neutrino-DM 
scattering cross section over DM mass. Those constraints
have been then compared with 
the constraint from observation of neutrinos from supernova 1987A. 

Paolo Serra, on the other hand, 
presented new results on the possible connection between
neutrino physics and dark energy.
In a recent analysis \cite{serra}, by combining the $^{76}$Ge \BBz result with
the WMAP 3-years data and other cosmological datasets a constraint on the
dark energy equation of state of $-1.67< w <-1.05$ has been obtained
at $95\%$ c.l.,  ruling out a cosmological constant. 
Interestingly enough, coupled neutrino-dark energy models may be 
consistent with such equation of state. 
While future data are certainly needed for a confirmation of 
the controversial Heildelberg-Moscow claim, our result 
shows that future laboratory searches for neutrino masses 
may play a crucial role in the determination of the dark energy properties. 

Properties of neutrinos, the lightest of all elementary particles, 
may be the origin of the entire matter-antimatter asymmetry of the
universe. This requires that neutrinos are Majorana particles, 
which are equal to their antiparticles, and that their masses 
are sufficiently small. Pasquale di Bari, in his talk, showed that 
leptogenesis, the theory explaining 
the cosmic matter-antimatter asymmetry, predicts that 
all neutrino masses are smaller than $0.2 eV$, which will be 
tested by forthcoming laboratory experiments and by cosmology
(see \cite{dibari}). 

Francesco Villante presented how nucleosynthesis restrictions 
on mixing of active neutrinos with possible sterile ones 
are obtained with the account of experimentally determined mixing 
between all active neutrinos (see \cite{villante}). 
The earlier derived bounds, valid 
in the absence of active-active mixing, have been reanalyzed and 
significant difference is found in the resonance case. The results 
are obtained both analytically and numerically by solution 
of complete system of integro-differential kinetic equations. 
A good agreement between analytical and numerical approaches 
seems demonstrated. 

Finally Tositaka Kajino (see \cite{kajino}) 
reminded us that neutrino oscillations 
affect light element synthesis through the neutrino-process 
in supernova explosions. The $^7Li$ and $^{11}B$ produced 
in a supernova explosion of a 16.2 solar-mass star model increase by 
factors of 1.9 and 1.3 in the case of large mixing angle solution 
with normal mass hierarchy and $\sin^{2}2\theta_{13} > 0.002$ 
compared with those without the oscillations. In the case of 
inverted mass hierarchy or nonadiabatic $13$-mixing resonance, 
the increment of their yields is much smaller. 
Neutrino oscillations raise the reaction rates of charged-current 
neutrino-process reactions in the region outside oxygen-rich layers. 
The number ratio of $^7Li/^{11}B$ 
could be a tracer of normal mass hierarchy and relatively large 
$\theta_{13}$, still satisfying $\sin^{2}2\theta_{13} < 0.1$, 
through future precise observations in stars having strong supernova component.

\section{Conclusions}
Neutrino nature and masses represent key items for neutrino Physics in the next years. Some of the most powerful tools to scrutinize this sector, cosmology, \BBz and $\beta$ decay, were discussed in the session. Cosmological bounds and \BBz measurements offer the best sensitivities but strongly depend on to some extent untested assumptions the first and on nuclear matrix  elements uncertainties the latter. Efforts to improve \BBz nuclear matrix elements reliability are in progress and their interplay with experimental activities will be crucial in the next future. On the other hand the rich program of experimental activities presented in the session is approaching the phase of production of results. 
Current experiments are probing the quasi-degenerate mass region, while a substantial coverage of the "inverted hierarchy" region would be achieved in the next generation \BBz experiments. 
These experiments will accommodate O(ton) of double beta decay emitters and will require improvements on background reduction and isotopes production. 
%The physics capability of these large-scale experiments has to be investigated in detail taking into account also the progress in the determination of mixing parameters in oscillation experiments and the progress in the knowledge of nuclear matrix elements. Different nuclear isotopes will be necessary to minimize the impact of uncertainties in matrix elements to the extracted mass values and different experimental techniques will help to establish the effect. 
Concerning the quasi-degenerate mass region, the latest WMAP results seem to not leave much room for direct mass measurements with 200-meV sensitivity. Moreover, cosmological sensitivity  is going to further improve with results from the Planck satellite mission.  However, despite the impressive success of precision cosmology, cosmological bounds are derived within a system of assumptions and interpretations. Considering the importance of the neutrino mass  question, and the difficulty in associating the cosmological limit to a precise systematic confidence level, direct measurements should be pursued up to their eventual technological limits. Probing this region is also crucial to verify the claim on \BBz of $^{76}$Ge, pointing to \amee value  around 0.4 eV.

In few words, several topics have been presented at our session.
All of them extremely stimulating and worthwile of future
developments and investigations. It will be duty of
future experiment, on earth, in space or ``in the universe''
to scrutinize all these important hypotheses.

{\it Acknowledgements}
We would like to thank the Organizers of the 
{\em NOW-2006\/} workshop.

%\end{document}

\end{document}